\documentclass[sigconf]{acmart}

\usepackage{comment}
\usepackage{float}
\usepackage{csquotes}
\usepackage{tabularx}
\usepackage{xcolor}
\usepackage{graphicx}
\usepackage{subcaption}
\usepackage{enumitem}
\usepackage{fvextra}

\BeforeBeginEnvironment{minted}{\vspace{-0.05cm}}
\AfterEndEnvironment{minted}{\vspace{-0.05cm}}

\newcommand{\hr}{\noindent\rule{\linewidth}{0.3pt}}
  
\newcommand{\circledigit}[1]{\textbf{\normalsize{\textsf{\textcircled{\footnotesize{#1}}}}}}

\newcommand{\eg}{\textit{e.g.}}
\newcommand{\ie}{\textit{i.e.}}

\newcommand{\etal}{\textit{et al.}}
\newcommand{\system}{\textsc{ReFinE}}
\newcommand{\proceedings}{\textsc{CHI \textquotesingle24}}

\newlength\myindent
\AtBeginDocument{%
  \providecommand\BibTeX{{%
    \normalfont B\kern-0.5em{\scshape i\kern-0.25em b}\kern-0.8em\TeX}}}

\copyrightyear{2026}
\acmYear{2026}
\setcopyright{cc}
\setcctype{by}
\acmConference[DIS '26]{Designing Interactive Systems Conference}{June 13--17, 2026}{Singapore, Singapore}
\acmBooktitle{Designing Interactive Systems Conference (DIS '26), June 13--17, 2026, Singapore, Singapore}
\acmDOI{10.1145/3800645.3812860}
\acmISBN{979-8-4007-2563-0/2026/06}

\begin{document}

\newcommand\todoit[1]{{\color{red}\{TODO: \textit{#1}\}}}
\newcommand\todocite{{\color{red}{CITE}}}

\definecolor{lightblue}{RGB}{212, 235, 255}
\definecolor{orange}{RGB}{255, 105, 0}
\definecolor{lightgreen}{RGB}{177, 231, 171}
\definecolor{lightyellow}{RGB}{255, 255, 148}

\newcolumntype{P}[1]{>{\centering\arraybackslash}p{#1}}
\newcolumntype{L}[1]{>{\raggedright\let\newline\\\arraybackslash\hspace{0pt}}m{#1}}
\newcolumntype{R}[1]{>{\raggedleft\arraybackslash}p{#1}}
\newcommand\tworows[1]{\multirow{2}{*}{\shortstack[l]{#1}}}
\newcommand\tworowsc[1]{\multirow{2}{*}{\shortstack[c]{#1}}}
\newcommand\threerows[1]{\multirow{3}{*}{\shortstack[l]{#1}}}

\title[\system{}]{\system{}: Streamlining UI Mockup Iteration with Research Findings}

\author{Donghoon Shin}
\orcid{0000-0001-9689-7841}
\email{dhoon@uw.edu}
\affiliation{%
  \institution{University of Washington}
  \city{Seattle}
  \state{WA}
  \country{USA}}

\author{Bingcan Guo}
\orcid{0000-0002-9001-8727}
\email{bguoac@uw.edu}
\affiliation{%
  \institution{University of Washington}
  \city{Seattle}
  \state{WA}
  \country{USA}}

\author{Jaewook Lee}
\orcid{0000-0002-1481-9290}
\email{jaewook4@cs.washington.edu}
\affiliation{%
  \institution{University of Washington}
  \city{Seattle}
  \state{WA}
  \country{USA}}
  
\author{Lucy Lu Wang}
\orcid{0000-0001-8752-6635}
\email{lucylw@uw.edu}
\affiliation{%
  \institution{University of Washington,\\Allen Institute for AI}
  \city{Seattle}
  \state{WA}
  \country{USA}}

\author{Gary Hsieh}
\orcid{0000-0002-9460-2568}
\email{garyhs@uw.edu}
\affiliation{%
  \institution{University of Washington}
  \city{Seattle}
  \state{WA}
  \country{USA}}
  
\renewcommand{\shortauthors}{Shin, et al.}

\begin{abstract}
Although HCI research papers offer valuable design insights, designers often struggle to apply them in design workflows due to difficulties in finding relevant literature, understanding technical jargon, the lack of contextualization, and limited actionability. To address these challenges, we present \system{}, a Figma plugin that supports real-time design iteration by surfacing contextualized insights from research papers. \system{} identifies and synthesizes design implications from HCI literature relevant to the mockup's design context, and tailors this research evidence to a specific design mockup by providing actionable visual guidance on how to update the mockup. To assess the system's effectiveness, we conducted a technical evaluation and a user study. Results show that \system{} effectively synthesizes and contextualizes design implications, reducing cognitive load and improving designers' ability to integrate research evidence into UI mockups. This work contributes to bridging the gap between research and design practice by presenting a tool for embedding scholarly insights into the UI design process.
\end{abstract}



\begin{teaserfigure}
  \includegraphics[width=\textwidth]{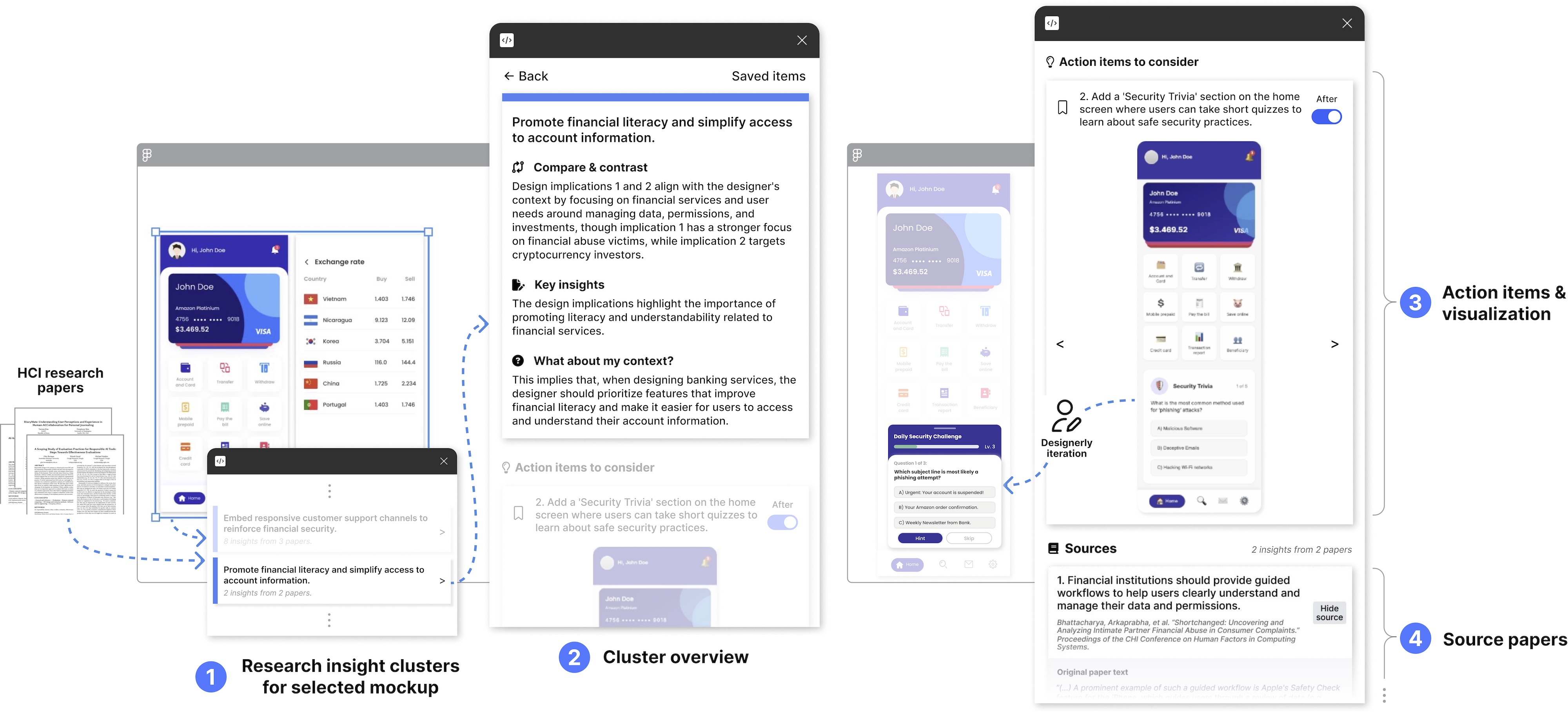}
  \caption{An overview of the \system{}-aided design workflow. Based on a user's selected UI mockup on the Figma canvas, the system generates \circledigit{1} clusters of research insight informed by HCI research papers. Designers can examine \circledigit{2} each cluster to understand the synthesized design implications before reviewing \circledigit{3} action items and their visualizations, which project how potential design edits could be made to the interface. Designers can use these to perform designerly iteration, modifying or adapting the proposed changes at their own discretion. They can also explore the \circledigit{4} source papers that informed each cluster.}
  \label{fig:keyscreen}
\end{teaserfigure}

\maketitle
 
\section{Introduction}

Scholarly papers in human-computer interaction (HCI) serve as an invaluable repository of emerging design implications. Guided by research findings, these implications are intended to support practical design decisions. Yet, design implications within these papers are often underutilized and fail to be effectively integrated into the workflows of design practitioners~\cite{colusso2017translational}. Prior research in translational science has identified several core challenges that impede the translation of scholarly knowledge into practical design action, including difficulties in finding relevant scholarly resources~\cite{colusso2017translational, bailis2016introducing},\footnote{While prior literature has used both \textit{discovery} and \textit{retrieval} to describe the process of finding relevant literature, for ease of reading, discovery and retrieval will be referred to for the rest of the paper as \textit{retrieval}.} adapting the communicated content and format to make the insights more understandable to designers~\cite{shin2025what, colusso2017translational, norman2010research}, and contextualizing design implications within the designer's action space~\cite{shin2024paper, colusso2017translational, norman2010research}. These challenges are particularly pronounced during design prototyping, when designers who often work under time constraints iterate rapidly on emerging design ideas and mockups, making it impractical to continuously retrieve, translate, and contextualize scholarly insights in real-time~\cite{ferreira2007agile,suleri2019ui}.

Recent work has explored making research papers more accessible to broader audiences. For instance, prior works in HCI have proposed several research prototypes (\eg{}, personalized paper alerts~\cite{lee2024paperweaver}, Q\&A systems~\cite{august2023paper}, recursive abstract exploration~\cite{fok2024qlarify}) to support the awareness and consumption of jargon-filled papers. However, these methods largely focus on addressing retrieval issues or summarizing paper insights, and do not directly address the needs of design practice. In the design domain, a handful of recent works have introduced tools to translate design implications into prescriptive formats (\eg{}, design cards~\cite{shin2024paper, nurain2024designing}), demonstrating potential to convey scholarly insights. Yet, these efforts have not addressed the retrieval problem or significantly improved the actionability of the research, which is especially critical during the prototyping stage, where identifying relevant insights and getting actionable recommendations are key to supporting iterations.

We introduce \system{} (\textbf{\textsc{Re}}search \textbf{\textsc{Fin}}dings for \textbf{\textsc{E}}vidence-informed UI design iteration), the first AI-powered design support system that facilitates real-time, in-situ UI mockup iteration by leveraging design implications from published research works. \system{} allows designers to draw inspiration from the latest research with minimal manual effort, by automatically retrieving, translating, and contextualizing design implications present in HCI research papers---directly within their existing design workspace (\ie{}, Figma). To achieve this, our work addresses key challenges in translational science: (i) eliciting \textit{design context} dimensions (\ie{}, target user, domain, modality, pain point, client, metric) from each paper that represent what designers perceive as `relevant' when consuming scholarly knowledge \cite{shin2025what}, and using these to retrieve papers, and identify and cluster design implications, (ii) generating summaries and translating scholarly insights by drawing analogies to bridge the gap between research findings and the design mockup, and (iii) providing `action items' for their design and visualizing how these changes could be applied on a designer's mockup reconstructed as an HTML.

\system{} leverages multimodal large language models (LLMs) for retrieval, mockup understanding, paper understanding, and generating actionable insights. To evaluate the effectiveness and accuracy of the system, we conducted a series of technical evaluations of intermediate model outputs on examples of mobile UI design iteration. Using a dataset constructed from research papers and mobile UI design mockup examples, we measured and reported the latency for generating each component in \system{}. Our comparative study evaluating \system{} components against alternative LLMs, input modalities, and techniques justifies our design choices. Additionally, we assessed the reliability of LLM-driven mockup understanding and HTML generation\footnote{LLMs have demonstrated a strong understanding of HTML~\cite{gur2023real, deng2022dom, li2021markuplm, burns2023suite, aghajanyan2021htlm}.} within our system (\eg{}, eliciting dimensions of the design context, generating action items), finding that the outputs were largely faithful to their original inputs.

We further conducted a within-subjects user study with professional designers and design students in UI/UX ($N=12$), where participants engaged in prototyping iterations both with and without \system{}. Our findings indicate that \system{} improved the communication of scholarly insights across key metrics~\cite{sas2014generating}, \ie{}, improving generativity, inspirability, actionability, validity, generalizability, and relevance, without undermining originality. Also, participants were able to create significantly more design edits and quickly reach design iterations on the canvas when using \system{}. \system{} was also easy to use, allowing designers to integrate complex scholarly findings with minimal cognitive load and effort. Qualitative interviews revealed several factors contributing to these improvements, highlighting that the visualization of action items served as a generative aid for design iterations, allowing designers to integrate scholarly evidence without disrupting their creative workflow. We also discuss potential future enhancements for supporting UI mockup iterations with research findings.

To summarize, we contribute the following:
\begin{itemize}
    \item \system{}, a system that facilitates UI mockup iterations by automatically inferring dimensions of the design context directly from the visual workspace to retrieve, translate, and contextualize scholarly insights;\footnote{The codebases for both (i) the frontend plugin and (ii) the backend server---including the prompt systems---are available here: \url{https://github.com/donghoon-io/refine}.}
    \item Results from an evaluation study, including technical evaluations and a user study, demonstrating the system's ability to enhance the design iteration process by making scholarly insights more accessible and usable;
    \item Insights into how evidence-informed design can be further facilitated through leveraging research findings in the design workflow.
\end{itemize}
\section{Related Work}

\subsection{Translational Science for Design}

In the field of HCI, much of the published scholarly literature provides design implications (also known as guidelines) that could potentially guide design iterations and improve designs. Yet, these resources often remain underutilized by design practitioners~\cite{colusso2017translational, norman2010research}. Previous research in translational science for design has pinpointed several obstacles that make it challenging to consume and leverage scholarly insights in design practice~\cite{shin2024paper, colusso2017translational, colusso2019translational}.

One major hurdle is finding and retrieving relevant resources~\cite{colusso2017translational, bailis2016introducing}. Designers often find it difficult to craft search queries that can identify papers relevant to their specific context~\cite{colusso2017translational, bailis2016introducing}. For example, a designer looking for UI patterns to boost accessibility might not know to search for terms like ‘inclusive design heuristics,’ leaving valuable design implications out of reach. This problem is worsened by existing paper search engines, which largely rely on basic metadata (\eg{}, titles, abstracts) for search (\eg{},~\cite{arxiv_search}), offering little depth or precision for design practitioners.

Another challenge is translating scholarly insights in a way that helps designers understand their use in design work~\cite{colusso2017translational, norman2010research}. Despite the HCI community's common practice to communicate design implications in published papers, designers often find these papers difficult to read~\cite{colusso2017translational}. Papers are often full of jargon, and the design implications may not be clearly communicated in a way that highlights insights that are generative, inspiring, and actionable ~\cite{sas2014generating}.

Lastly, scholarly design implications are not sufficiently contextualized within a designer's specific action space~\cite{shin2024paper, norman2010research}. This problem is exacerbated by the fact that research papers themselves rarely match a designer's project's exact constraints or goals, such as specific target users or platforms~\cite{colusso2017translational}. While past work suggests even out-of-domain studies can offer valuable insights to support individual designers’ needs~\cite{shin2025what}, designers often lack the know-how or tools to reinterpret these findings effectively for their unique situation, leading them to perceive the research as less useful or its implications as too generalized~\cite{colusso2017translational}. Additionally, even when designers dig into relevant papers and grasp the concepts, spelling out the practical and actual next actions to take within their own working design remains difficult. For instance, insights from a paper on online learning for older adults might hold value but fail to offer direct, actionable guidance for a designer working on a mobile design for seniors. In fact, prior research on auto-generated tools demonstrated that transforming scholarly insights into concise formats, while helpful, was insufficient to be fully supportive of the ‘what should I actually do in my design?’ question~\cite{shin2024paper, colusso2017translational}.

To tackle these key challenges in translational science for design, we present a systematic approach that weaves HCI research findings directly into the iterative UI design process. Our tool simplifies resource discovery with context-aware paper retrieval tailored to designers' needs through design context extraction, supports the translation of insights for specific design scenarios, and serves up practical, tailored recommendations for immediate integration. As such, we aim to bridge the gap between scholarly knowledge and design practice, making HCI research a more useful asset for everyday design work.

\subsection{AI-supported UI Design Iterations}

The process of refining UI designs based on feedback, testing, or newly added insights~\cite{adams2002understanding, nielsen2002iterative} is critical throughout the UI design lifecycle---from early-stage wireframing to high-fidelity prototyping~\cite{ferreira2007agile}. Designers continuously refine interfaces to enhance usability and creative expression, ensuring alignment with user needs and design principles, while validating design choices and incorporating new knowledge effectively. AI has been supporting this process by automating the mundane tasks to allow designers to concentrate on mentally demanding and creative work.
Recent advancements in AI, especially in LLMs, are providing designers with useful methods to approach the iterative process, enabling innovative ways to generate ideas, assess designs, and incorporate insights.

One key area where AI provides support for UI design iterations is in exploring inspirational UI examples. While designers traditionally relied on manually searching through image repositories~\cite{wu2021exploring}, AI supports are being increasingly used to enhance semantic search capabilities to facilitate fast and relevant queries. For instance, building on foundations laid by large-scale UI datasets (\eg{},~\cite{deka2017rico, leiva2020enrico}), systems like VINS~\cite{bunian2021vins} enabled visual search by leveraging embedded screen segments. Another work explored the use of multimodal LLMs to capture deeper semantic categories even without the screen-associated app metadata for more refined search results~\cite{park2025leveraging}. However, merely providing inspiration might be limited in that designers often seek supporting evidence behind AI recommendations to substantiate them~\cite{zhong2024ai}---a gap that these inspiration-focused approaches may not fully address. Furthermore, adapting retrieved examples effectively to specific design contexts remains a challenge~\cite{herring2009getting}.

Beyond inspirational support, AI is increasingly being applied to provide more contextualized assistance within the current design mockup. For instance, LLMs have begun to assist designers in understanding their mockups, such as by facilitating conversational interactions with UI designs~\cite{wang2023enabling}. Not limited to understanding UI, prior work has also proposed more direct forms of support, such as interfacing UI design iteration with code development loop~\cite{ma2024didup} and enhancing UI iterations by blending example design components~\cite{lu2024misty} using LLMs. While many of these systems operate in specialized, custom-built environments, a handful of efforts have extended such capabilities to widely adopted platforms like Figma---as seen in tools that automate usability evaluation using heuristics~\cite{duan2024generating}, though their application remains limited in providing textual feedback using a predefined set of design guidelines.

Building on this body of research, in this work, we propose a novel approach that orchestrates the rich knowledge contained within and retrieved from published research papers and AI to support UI design iterations. Specifically, we demonstrate using the generative and understanding capabilities of multimodal LLMs to retrieve relevant design insights from scholarly literature, translate these findings into actionable suggestions, and contextualize them within a designer's specific ongoing design iteration. We posit that this method addresses the need for scholarly research-backed guidance---moving beyond generic heuristics or serendipitous inspiration---while significantly reducing the effort required to discover and apply these scholarly insights, integrating seamlessly into designers' workspace.
\section{The R{\small e}F{\small in}E System}



\begin{figure*}[t!]
    \centering
    \includegraphics[width=\linewidth]{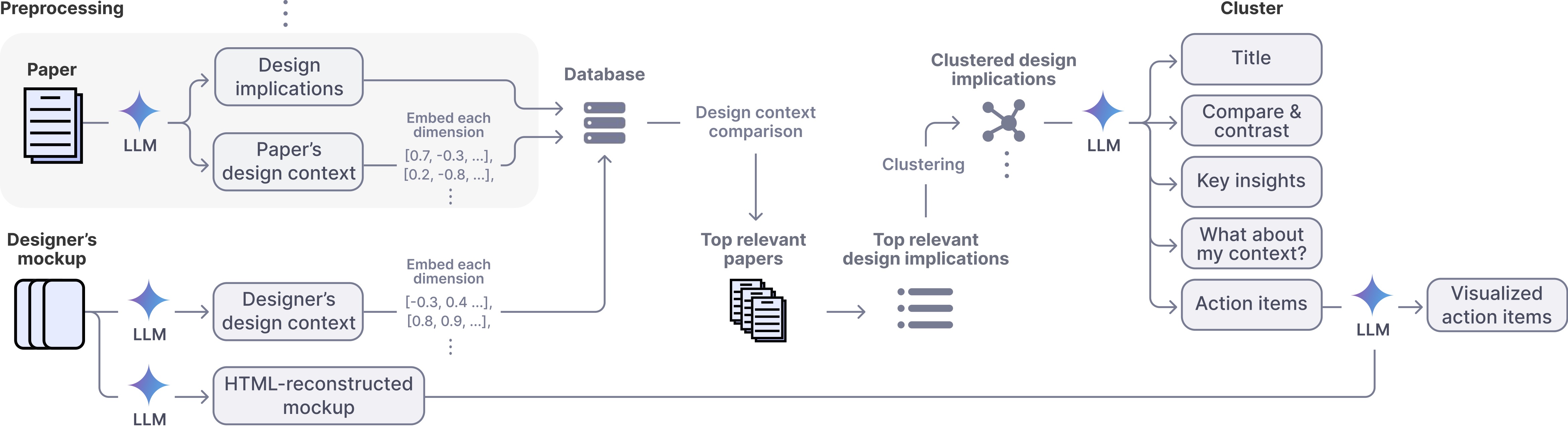}
    \caption{Pipeline for generating the components provided by \system{}}
    \label{fig:pipeline}
\end{figure*}

\system{} (see~\autoref{fig:keyscreen}) is an in-situ design support system powered by design implications from HCI research papers. Implemented as a Figma plugin, it interprets a user's current mockup to infer its context, then retrieves, synthesizes, and translates relevant research evidence into actionable UI suggestions tailored to that design.

Users begin by selecting a design mockup within Figma. \system{} analyzes the selection to automatically extract its design context dimensions (\eg{}, domain, target user, modality), which users can review and confirm. The system then uses this context to retrieve the most relevant scholarly papers from a preprocessed repository, from which it identifies pertinent design implications and organizes them into coherent insight clusters.

For each cluster, \system{} provides a summary and context-specific insights tailored to the selected mockup, along with action items that illustrate potential research-informed design changes. All insights are linked to their source papers. Below, we describe the design and technical considerations of each \system{} component.

\subsection{Preprocessing}

\subsubsection{Building the paper index}

To facilitate the retrieval of design recommendations, we focus on indexing papers along key dimensions of their design context that influence how designers assess research relevance. Prior work~\cite{shin2025what} identified six such dimensions: (i) target user, (ii) domain (\ie{}, the industry or field the design targets), (iii) modality (\ie{}, the primary interaction mode or medium), (iv) pain point (\ie{}, the main user problem to address), (v) client (\ie{}, the entity or stakeholder commissioning or benefiting from the design), and (vi) metric (\ie{}, the key performance indicator (KPI) used to measure success). We use these dimensions to index papers in \system{} and as the primary retrieval mechanism.

Starting from raw paper PDFs, \system{} converts each file into structured XML using GROBID~\cite{lopez2009grobid}, a machine learning-based paper parser. An LLM then extracts the six design context dimensions from the XML. Because not all papers contain all dimensions, the model is instructed to abstain when a dimension is absent. The extracted text for each dimension is then converted into vector representations using a text embedding model (Google's text-embedding-004~\cite{embedding}); missing dimensions are assigned an empty-string embedding (\ie{}, \texttt{“”}), and papers for which no dimensions can be identified are excluded from the retrieval index. All embeddings are stored as vector arrays in cloud storage.

In this work, we use an index of the exhaustive set of papers ($N=1{,}060$) from the ACM \proceedings{} proceedings. As a broad HCI conference spanning many subfields~\cite{chi2024_paper}, CHI provides coverage across diverse domains and supports topical generalizability. The index can also be constructed from other proceedings or curated paper collections and easily substituted into the system.

\subsubsection{Identifying design implications in papers}

Although HCI papers often include explicit design implication or guideline sections, many convey such insights implicitly, making a rule-based approach challenging. In this regard, Sas \etal{}~\cite{sas2014generating} defined a ruleset characterizing design implications, including their functions, taxonomy, sources, and heuristics. Motivated by prior work showing that rulesets and heuristics can effectively identify relevant entities in documents (\eg{},~\cite{fok2024qlarify}), we use this ruleset as instructions for an LLM to identify design implications.

Using each paper's XML representation, \system{} prompts an LLM with the definition of design implications from~\cite{sas2014generating} to extract an array of implications, each consisting of (i) the design implication text, (ii) its source paragraph, and (iii) a rationale for the model's derivation. When an implication is complete and clearly stated, the model uses it as-is; otherwise, it may make minor edits to ensure the implication is self-contained and preserves its original meaning. Because not all HCI papers include design implications~\cite{sas2014generating}, the model is instructed to return an empty array when none are identified. Across all papers, the LLM extracted at least one design implication from 81.3\% of them.

\subsection{Design Mockup Understanding \& Evidence Retrieval}

\subsubsection{Understanding the design mockup}\label{sec:mockup-querying}

To identify papers relevant to a designer's design mockup, \system{} extracts the same six design context dimensions directly from the selected mockup on the Figma canvas, following an approach analogous to paper preprocessing. Once the relevant mockup screen(s) are selected, \system{} converts them into a base64-encoded PNG and uses the image as input to prompt the model to extract the mockup's design context dimensions. Each extracted dimension is then embedded using the same text embedding model as for the papers.

\subsubsection{Retrieving papers relevant to the current design}\label{sec:retrieval}

To identify papers relevant to a design mockup, \system{} compares the mockup's dimensions with those of papers stored in the vector repository. The system first computes a summed embedding over all present dimensions for the mockup and compares it against the summed embedding of the corresponding dimensions for each paper in the repository. Papers are ranked by cosine similarity between these aggregated embeddings (see Appendix~\ref{app:algorithms} for the detailed mathematical formulation of the retrieval logic). To reduce information overload for downstream tasks, \system{} retains the top eight most similar papers, though this setting is configurable. All design implications from the retained papers are used in subsequent steps.

\subsubsection{Clustering the design implications}

\system{} then clusters the retrieved design implications into themes, enabling the generation of coherent groups of design insights. This approach eliminates the need to analyze each implication individually while enhancing validity by synthesizing insights from multiple sources~\cite{sas2014generating}. To achieve this, we employed hierarchical clustering\footnote{For hierarchical clustering, we use average linkage and cosine distance as the distance metric.} on the embeddings of design implications. The number of clusters is determined by evaluating candidate cluster counts using the mean silhouette score~\cite{rousseeuw1987silhouettes} and choosing the configuration that maximizes this metric (see Appendix~\ref{app:algorithms}). In our pilot study with 20 UI mockups, the median optimal cluster count was 6, and no case exceeded 10 clusters; accordingly, we capped the cluster count at 10 for computational efficiency. The resulting clusters group related design implications, and each cluster's output includes the original implications (paper ID, text, and source paragraph), forming the basis for the translated insights in \S\ref{sec:translation}.

\subsection{Translating Insights based on Designers' Design Context}\label{sec:translation}

\subsubsection{Identifying similarities and differences between design implications and designer context}

Previous research has highlighted the importance of understanding both the similarities and differences between the context of the papers and the designer's design context to enhance comprehension of how the implications might be applied~\cite{shin2025what}. To support this, \system{} provides an overview of similarities and differences (\ie{}, \textit{Compare \& contrast}) between the designer's context and the design implications of the cluster.

Inspired by the literature demonstrating the effectiveness of LLMs in comparing and contrasting multiple entities (\eg{},~\cite{webb2023emergent, yu2023thought}), we leverage an LLM to identify and describe key similarities and differences. The model is first given a list of design implications, their source paragraphs, and the design context of each paper. Then, it is instructed to analyze similarities and differences in relation to the designer's design context, and the resulting summary of these similarities and differences is presented to the designer.

\begin{figure}[t!]
    \centering
    \includegraphics[width=\linewidth]{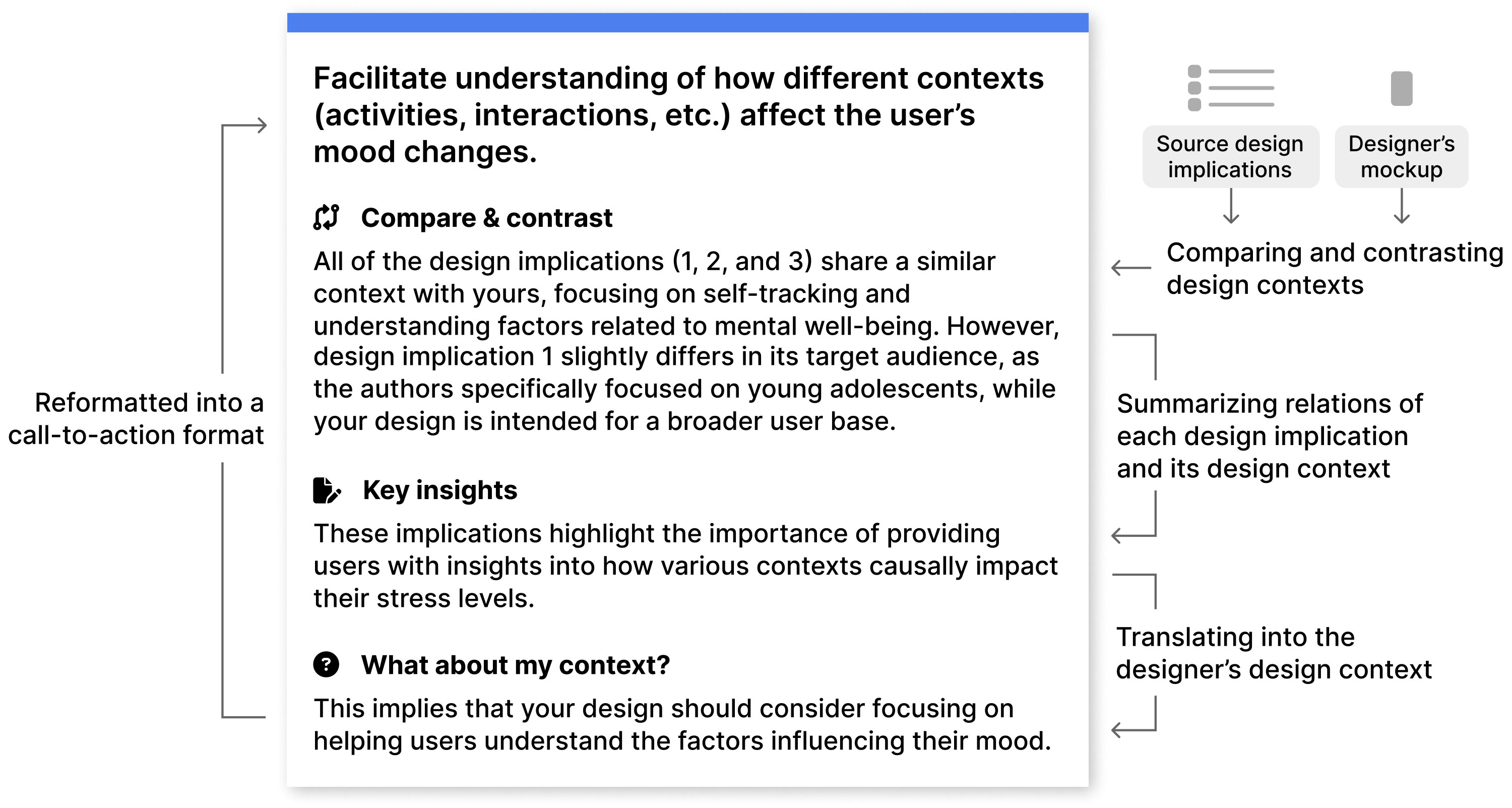}
    \caption{Contents of the cluster and their formulation}
    \label{fig:cluster-detail}
\end{figure}

\subsubsection{Tailoring design insights through analogy generation}

Because the design context underlying a paper's design implications rarely aligns perfectly with a designer's specific situation, these insights must be translated to fit individual needs. To bridge this gap, \system{} leverages analogical ideation~\cite{yu2014searching, gilon2018analogy}, a widely used approach for adapting design insights across contexts. Analogical ideation emphasizes relationships between surface-level information~\cite{gilon2018analogy}---in this case, the relationship between a design implication and its original design context---and has been extensively applied to facilitate the transfer of insights across domains~\cite{gilon2018analogy, yu2014searching, chan2011benefits, gentner1997structure}.

First, the LLM is prompted to define the relation~\cite{shin2025what, yu2014searching} between each design implication in a cluster and its source paper's design context using Chain-of-Thought prompting~\cite{Wei2022ChainOT}. The model then summarizes these relations, the result of which serves as the cluster summary (\ie{}, \textit{Key insights}). Here, if the cluster contains conflicting implications, the model is instructed to resolve the tension by prioritizing the one closer to the designer’s extracted design context, ensuring the tailored insight remains immediately actionable. This relational summary is then applied to the designer's specific situation to generate a tailored design insight (\ie{}, \textit{What about my context?}), derived from both the clustered design implications and their original contexts. Finally, the model transforms this tailored insight into a call-to-action, producing a catchy title (\ie{}, \textit{Cluster title}) that designers can browse from the cluster list.

\subsection{UI Mockup Reconstruction \& Action Item Generation}

In addition to supporting conceptual translation, \system{} also supports visual instantiation through contextualizing insights---by generating concrete action items and visualizing them within a designer's design mockup.

\subsubsection{Generating action items}\label{sec:action-item}

Based on insights tailored to the design context, \system{} generates up to three actionable suggestions situated within the designer's mockup (\ie{}, \textit{Action items}). An LLM is provided with an image representation of the designer's mockup, the cluster's implications, and the translated insights. It is then instructed to return action items, along with which mockup screen(s) each action item can be applied to. Additionally, recognizing that some changes (\eg{}, animations) cannot be easily visualized, the model returns a boolean flag indicating if the item is \textit{visually representable}---indicating if the item (i) applies directly to visual elements on the current screen and (ii) does not require generating new screens. If met, \system{} generates a visual preview as in \S\ref{sec:preview-application}; otherwise, it presents the action item as text-only.

\subsubsection{Constructing a preview of the mockup with the action item applied}\label{sec:preview-application}

To visualize these items, \system{} converts the design into manipulable HTML, applies specific changes, and renders the result:

\paragraph{Reconstructing mockups}
\begin{figure}
    \centering
    \includegraphics[width=\linewidth]{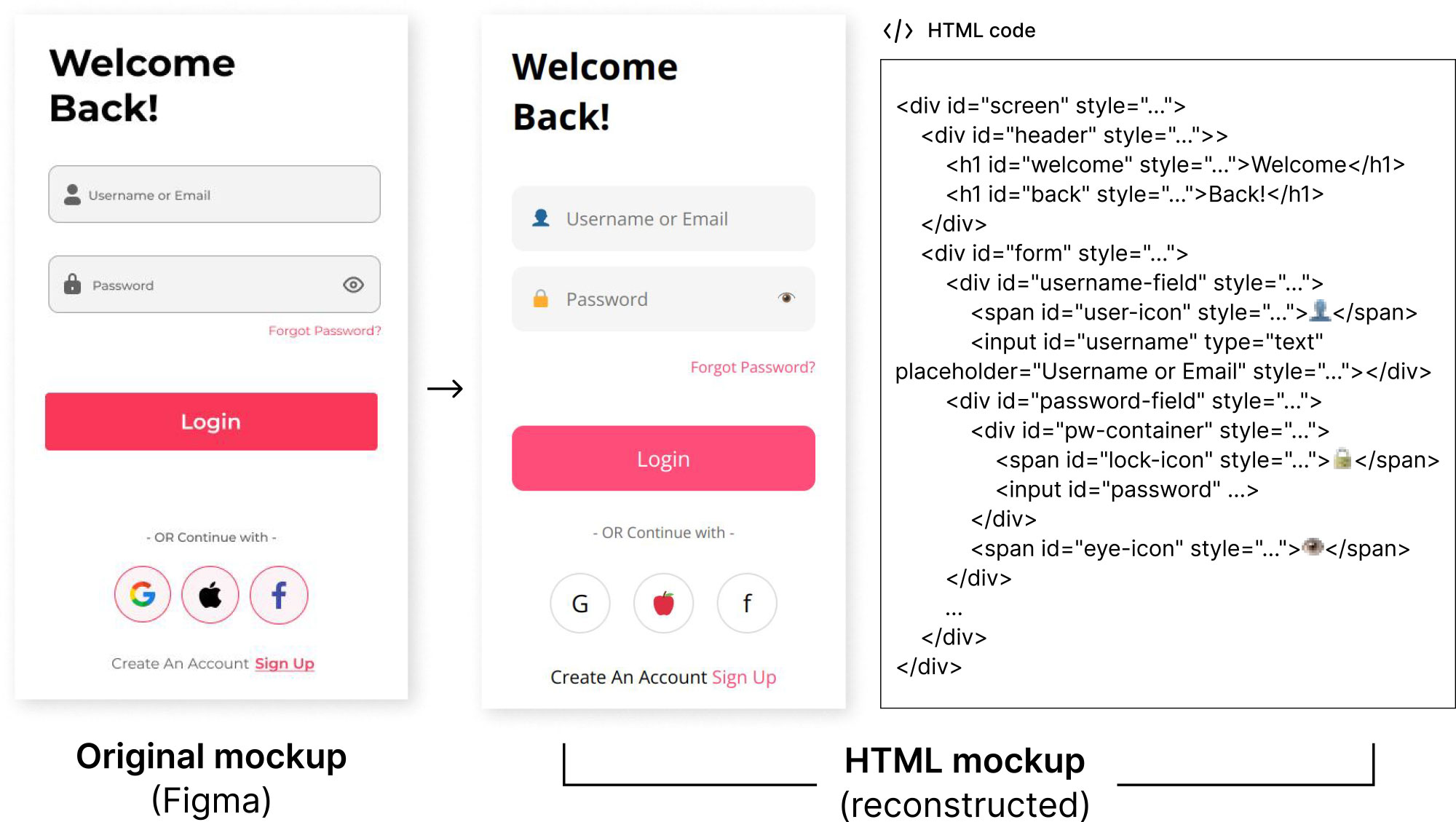}
    \caption{An example of mockup reconstruction in our pipeline. Although the reconstructed HTML may not perfectly match the original mockup, it faithfully mirrors the components in the original mockup and allows for visualization of action items.}
    \label{fig:mockup_reconstruction}
\end{figure}
Since programmatic manipulation of Figma's design mockups is limited,\footnote{https://www.figma.com/plugin-docs/} we decided to have the system reconstruct the mockup in HTML, a format known to be effectively handled by LLMs~\cite{gur2023real, deng2022dom, burns2023suite}. Concurrently with cluster generation, an LLM analyzes the mockup image to generate its HTML/CSS representation. To reduce latency, the model is prompted to return simplified placeholder stand-ins (\eg{}, emojis, colored <div/> containers) rather than pixel-perfect assets in cases where faithful reconstruction would be prohibitively complex (\eg{}, icons), preserving the semantic structure and layout while maintaining computational efficiency.

\paragraph{Applying action items}\label{sec:system:action-items}

When applying an action item to the HTML, \system{} avoids the computationally expensive process of fully regenerating the mockup by adopting an `edit-only' approach. The model is instructed to identify targeted DOM operations---\ie{}, \textit{add}, \textit{remove}, or \textit{replace}---that specify both the targeting element IDs and the modifications required to implement the action item described in \S\ref{sec:action-item} (\eg{}, see the instruction below for a \textit{replace} operation).

\begin{figure}[t!]
    \centering
    \includegraphics[width=.75\linewidth]{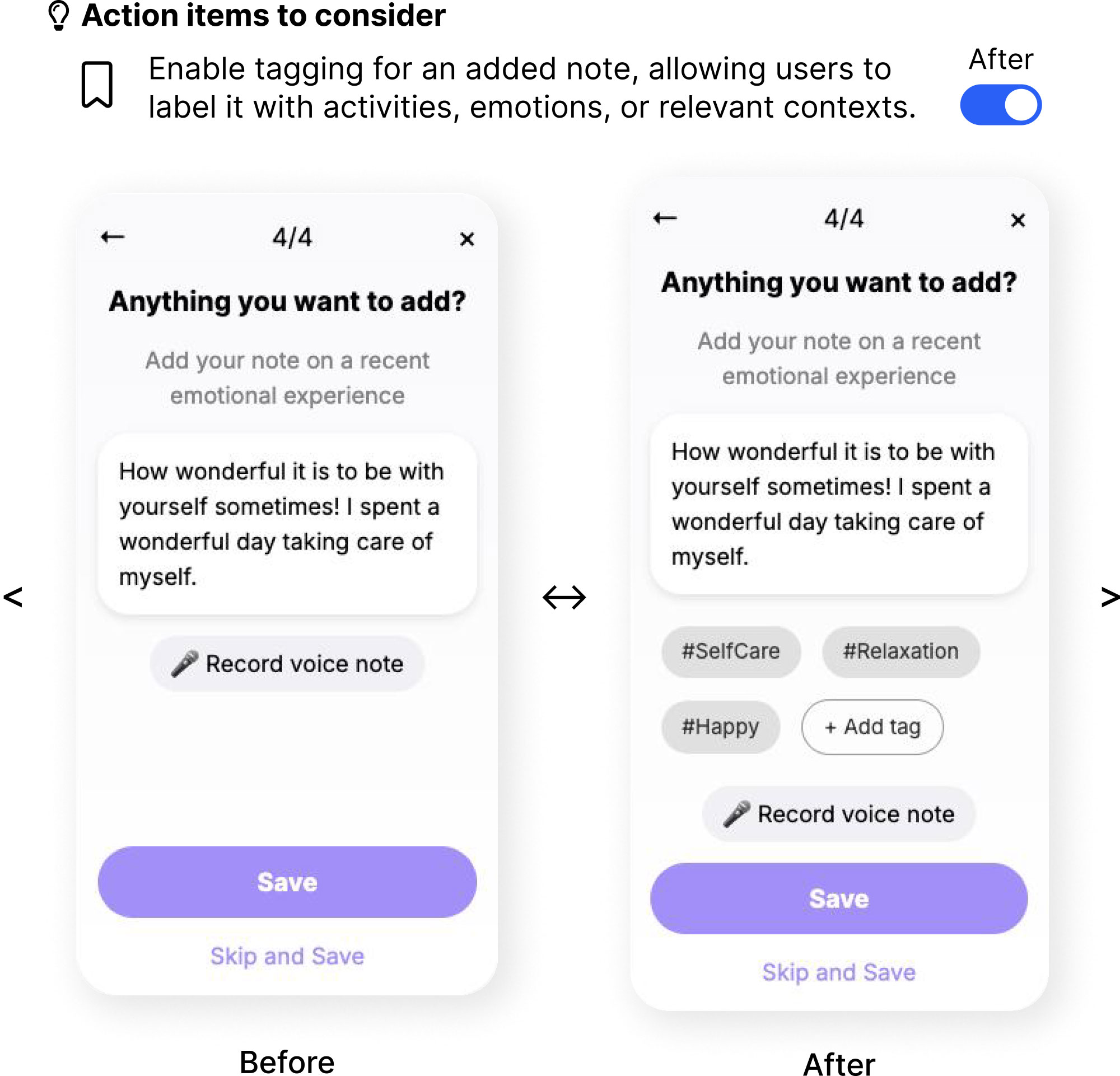}
    \caption{An example of an action item and its visualization. With the reconstructed HTML, \system{} selects the appropriate screen(s) from the design mockup and visualizes how each action item can be applied. Users can toggle between the before and after visualization, assisting designers in grasping the differences. Users can also bookmark each action item.}
    \label{fig:action-item}
\end{figure}

\hr{}
\texttt{...\newline}
\texttt{\footnotesize{[Rules (for \textquotesingle replace\textquotesingle)]}}
\begin{itemize}[label=-,leftmargin=.51cm, topsep=0pt]
    \footnotesize
    \item \texttt{When a certain element should be modified, target the smallest possible (lower-level) DOM elements to avoid redundant updates.}
    \item \texttt{Do not modify parent or sibling elements unless absolutely required for the change to work.}
    \item \texttt{\{style-related instructions\}}
\end{itemize}\vspace{.2cm}
\texttt{\footnotesize{[Output format]}}
\begin{Verbatim}[fontsize=\footnotesize,breaklines,breaksymbolleft={},framesep=1mm,xleftmargin=0pt,tabsize=4]
{     
    "reference_element_id": String, // an id for the element that needs to be changed
    "edited_element": String, // an HTML code for the modified element
    "rationale": String
}
\end{Verbatim}
\texttt{...\newline}
\hr{}
\system{} then parses the original HTML, applies these targeted edits, and renders the updated view (see \autoref{fig:action-item}), displaying quick visual inspiration without the need for full code regeneration.

\subsection{Implementation}

The \system{} system consists of two main components: (i) a backend server and (ii) a frontend web app. The backend server operates on Python 3.11 using Google Cloud's Ubuntu infrastructure and handles the backend operations, including LLM computations, clustering, and evidence retrieval. The frontend web app is built with SvelteKit,\footnote{https://svelte.dev/} a JavaScript-based web framework, and is integrated with Figma's Plugin API to function as a Figma plugin and enable interaction between the Figma workspace and the plugin. For parsing HTML code, we used the BeautifulSoup4\footnote{https://www.crummy.com/software/BeautifulSoup/} library.


For natural language and vision-language tasks, we provide a detailed description of our model selection in the following section based on our technical evaluation, along with the input modality used to reconstruct the mockups. While it justifies specific configurations, \system{}'s modular architecture enables its multimodal LLM to be easily replaced with any alternative model.

\begin{figure*}[t!]
    \centering
    \includegraphics[width=\linewidth]{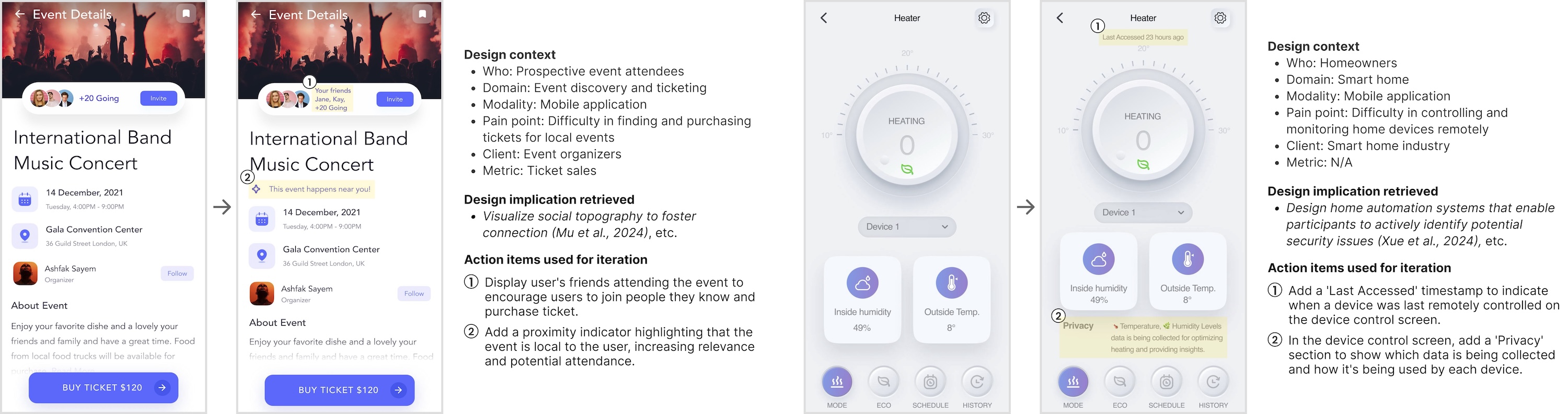}
    \caption{Design outputs from iterating two example mockups with \system{}-generated action items. For each example, we present the extracted design context, an example of retrieved design implications, and action items used for the iteration. For conciseness, a single screen pair (before/after) is shown for each mockup, and edits are highlighted with a yellow background.}
    \Description{}
    \label{fig:proof-by-demonstration}
\end{figure*}
\section{Technical Evaluation Study}

\system{} is designed to balance the key technical attributes: efficiency (\ie{}, ensuring low latency for real-time support) and fidelity (\ie{}, ensuring visual reconstructions are accurate enough to be credible), while ensuring that the generated components are relevant (\ie{}, ensuring the system accurately interprets design contexts to generate applicable insights). Our technical evaluation assesses the performance of pipeline components along these attributes.

\subsection{Datasets}

To ensure our evaluation reflects real-world design scenarios, we constructed datasets of UI mockups and HCI papers. For the UI mockup dataset, the research team first reviewed Figma's publicly accessible mobile app design mockup templates.\footnote{https://www.figma.com/community/mobile-apps} Starting with the top designs, the team manually screened and excluded those that were (i) too thematically similar to earlier designs or (ii) primarily layout-focused without meaningful semantic content. We continued this process until we reached a total of 50 mockups for inclusion, representing diverse themes. Each mockup consisted of 3 to 5 screens. For the HCI paper dataset, we used 50 randomly selected full papers from the \proceedings{} conference proceedings~\cite{chi2024} to evaluate the system's ability to extract context and retrieve relevant insights.

\subsection{Optimizing UI Reconstruction and Visualization}\label{sec:design-justification}

To visualize research-informed action items directly on the user's design, \system{} converts the static Figma mockup into manipulable HTML/CSS. This reconstruction step creates a technical trade-off: the generated HTML must possess high visual fidelity to the original mockup to ensure the designer recognizes the interface as their own, while maintaining efficiency to prevent disrupting the creative workflow.

To assess this trade-off, we evaluated architectural alternatives along these two competing attributes. For efficiency, we measured the total time required for generation. For fidelity, we measured visual similarity, defined as the cosine similarity between the visual transformer embeddings (vit\_base\_patch16\_224~\cite{dosovitskiy2020image}) of the original Figma screenshot and the rendered image of the reconstructed HTML. By probing these metrics, we sought to justify design choices that reliably replicate the visual structure for algorithmic manipulation, without introducing bottlenecks.

\subsubsection{Base model selection}
We compared three flagship commercial multimodal LLMs for HTML reconstruction (see \autoref{tab:model-comparison}): Gemini 2.0 Flash~\cite{gemini}, Claude 3.5 Sonnet~\cite{claude}, and GPT-4o~\cite{gpt4o}. Our results showed that, while Claude 3.5 Sonnet achieved marginally higher visual similarity ($0.7963 \pm 0.1590$) than Gemini 2.0 Flash, its latency was considerably higher ($17.006s \pm 4.914$). Thus, we selected Gemini 2.0 Flash as it offered the best trade-off, achieving comparable visual similarity ($0.7875 \pm 0.1437$) with significantly lower latency ($10.794s \pm 3.021$). Given that HTML reconstruction is the primary latency bottleneck in our pipeline involving both text and visual understanding, we extended this model choice to the other components to maintain architectural consistency.

\subsubsection{Input modality}
We investigated which input source yields the best reconstruction quality relative to latency. We compared providing the model with the mockup image alone, the JSON representation (provided by the Figma API) alone, or a combination of both (see \autoref{tab:modality-comparison}). Interestingly, providing JSON data along with an image resulted in lower visual similarity and increased latency compared to using the image alone. Using the image alone resulted in the highest visual similarity ($0.7875 \pm 0.1437$), achieved with low latency ($10.794s \pm 3.021$). Providing JSON alone performed the worst, where we hypothesize that the model overfitted to structural code cues rather than the visual rendering. As such, we decided to have \system{} rely exclusively on image inputs.

\subsubsection{Visualization technique}
To apply action items to reconstructed mockups, we proposed an `edit-only' approach, in which the LLM outputs only the necessary DOM modifications for computation efficiency when visualizing an action item (\S\ref{sec:system:action-items}). To evaluate how this approach affects the trade-off between efficiency and fidelity, we compared it with generating full HTML files for the same action items. We evaluated the application of 50 randomly selected action items ($N=150$ total operations; see \autoref{tab:preview-comparison}) to their corresponding reconstructed mockup screens. The edit-only method significantly reduced processing time ($t=26.34, p<.001$) from $10.25s$ to $2.91s$ per item, achieved without compromising accuracy (edit only: 95.3\% vs. full regeneration: 96.0\%). These results demonstrate that the proposed edit-only method substantially improves efficiency without sacrificing fidelity.

\subsection{Validating Extraction and Generation Components}

\subsubsection{Design context extraction}\label{sec:context_extraction}

The effectiveness of \system{} hinges on its ability to accurately interpret design contexts from both research papers and user mockups. To assess this, we evaluated whether \system{} consistently extracts relevant design contexts from these two sources. Two annotators, each holding professional UI/UX design degrees and more than three years of experience, independently reviewed the system's outputs using a custom Streamlit tool. Extraction accuracy was measured for six design context dimensions across 50 papers and 50 mockups, yielding 300 evaluations in total per each.

\paragraph{Papers} 
The system achieved 95.7\% accuracy in extracting design context dimensions from papers. The few identified errors include misclassification ($N=7$; \eg{}, labeling `online survey' as a modality when it was a research method), scope misalignment ($N=4$; \eg{}, specifying the target user as `users of LLMs' when the system was designed for any individual), and focus misinterpretation ($N=2$; \eg{}, treating `incidental learning' as a primary research domain when it was a study finding). The inter-rater agreement was strong, with a Cohen’s $\kappa$ of 0.82.

\paragraph{Mockups}
Similarly, the system achieved 94.3\% accuracy in extracting context from mockups. Identified errors include overgeneralization ($N=5$; \eg{}, categorizing as `publisher' without specifying the type of media), misclassification ($N=4$; \eg{}, identifying `touch' as a modality instead of `mobile app'), and incorrect metric identification ($N=8$; \eg{}, extracting metrics that were not explicitly signaled in the design). Strong inter-rater agreement was observed ($\kappa=0.88$).

\subsubsection{Relevance of generated insights}

Additionally, generated design guidance must remain consistent with its parent sources (\eg{}, original design implications, cluster titles, and designer mockups) to ensure validity. To assess this, annotators were first asked to evaluate the relevance of 50 randomly selected action items to the source design implications within their clusters on a 5-point Likert scale. The action items achieved a high average rating of 3.82/5, with only 14.0\% rated as slightly relevant or lower ($\kappa=0.81$; see \autoref{tab:relevance}).

We also evaluated whether the generated action items were contextually relevant to each cluster's title and whether \system{} correctly identified the relevant screen(s) in the mockup for applying them. Annotators found that 96.0\% of action items were relevant to their cluster titles ($\kappa = 0.79$) and 98.0\% accurately targeted the appropriate screens ($\kappa = 0.74$), demonstrating that \system{} produces actionable design insights that are both relevant and well-situated.

\subsection{Overall Latency Analysis}\label{sec:latency-analysis}

With the justified design choices and our dataset, we report the final latency for every component of \system{} (see \autoref{tab:latency-analysis}). For components involving multiple computations in parallel or outputs from distinct units (\eg{}, mockup reconstruction, generating cluster contents), we measured the time spent on each individual computation. The latency analysis for visualizing action items is based on the dataset used in \S\ref{sec:design-justification}.

This represents the performance of our optimized pipeline using our design choices, which achieved the best balance of efficiency and fidelity in our technical evaluations. While these benchmarks provide a technical baseline, we further evaluate the user perception of these latencies through our user study (\S\ref{sec:user-study}).


\begin{table}[ht!]
    \caption{Overall system latency analysis (metrics are computed per task \(\pm\) stdev, *: per unit; otherwise per design)}
    \footnotesize
    \begin{tabular}{lc}
    \toprule
    \textbf{System step} & \textbf{Latency (seconds)} \\
    \midrule
    Eliciting design context dimensions & \(3.167 \pm 1.019\) \\
    Mockup reconstruction* & \(10.794 \pm 3.021\) \\
    Retrieval \& clustering & \(2.004 \pm 0.400\) \\
    Generating translated insights* & \(2.982 \pm 0.326\) \\
    Generating action items & \(4.435 \pm 0.480\) \\
    Visualizing action item & \(2.913 \pm 1.215\) \\
    \bottomrule
    \end{tabular}
    \label{tab:latency-analysis}
    \Description{}
\end{table}
\section{User Study}\label{sec:user-study}

Our technical evaluation justified \system{}'s technical choices, evaluated the relevance of the generated outputs, and presented an overview of the performance. To further understand how \system{} is perceived and utilized in actual design iteration, we conducted a within-subjects user study involving 12 designers.

\subsection{Recruitment}

We began by recruiting designers from three design-focused university communities and one online community. Our recruitment post specified that participants must either (i) be currently working as professional UI/UX designers or (ii) be students pursuing a professional UI/UX design degree, who have experience using Figma in design work. As a result, we recruited 12 participants; of all, 8 identified as female, 4 male, and the average age was 26.0 (\(SD=4.5\)). The average length of their design experience was 4.2 years ($SD=2.3$).

\subsection{Procedure}

Each study was conducted remotely via Zoom. Before starting, each participant was asked to make sure the Figma desktop app was installed on their device. Also, we retrieved two publicly available mobile UI mockups from the Figma mobile app library---one for a financial banking app and the other for a travel app (see Appendix~\ref{apdx:study-details-user-study}), each containing four key screens.

Once the user joined the study, we introduced the objectives of our study. Then, each participant went through two design conditions, with order randomized to minimize any ordering effects: (i) engaging with scholarly insights and iterating on the mockup using \system{}-supported design (\ie{}, \system{} condition), and (ii) the baseline condition, which replicated the standard practice of evidence-informed design. In this baseline, participants searched the ACM Digital Library and synthesized knowledge to guide their iterations, while being allowed to freely use any familiar auxiliary tools to assist with paper consumption. Here, two participants used ChatGPT by copying and pasting paper content to ask questions about it. This condition served to establish an ecological baseline for the friction involved in discovering and translating scholarly work.

Each condition was randomly assigned one of the two UI mockups we prepared. To ensure a controlled comparison with our current implementation that leverages the \proceedings{} proceedings, we restricted the papers available for retrieval in both conditions to those published in \proceedings{}. Prior to each condition, we instructed participants on how to import the \system{} plugin into Figma, or access the ACM DL to search and retrieve full-text papers, based on the condition.

Each condition lasted 30 minutes. During the first 25 minutes, designers were asked to inform their designs with insights from research papers, and we encouraged them to iterate on the design without focusing heavily on the trivial visual aesthetics of their edits. During each condition, we notified them every 5 minutes to help them manage their time. Once the participants completed the design condition, they proceeded to a survey and filled out questionnaires for 5 minutes.

After completing both conditions, participants took part in a qualitative interview session, where they answered questions about their preferences and discussed the strengths, weaknesses, and potential enhancements for both components of \system{} and the overall system. The interview lasted approximately 20 minutes.

In total, the entire process lasted approximately 90 minutes. Each participant was compensated with a 50 USD gift card for their participation. The study protocol was reviewed and approved by the university's IRB.

\subsection{Measure \& Analysis}\label{sec:measure}

\subsubsection{Survey measures}

First, by utilizing \system{}'s design support, we hypothesized that designers could significantly lessen the effort required to extract scholarly insights from scholarly repositories. To evaluate the cognitive workload of participants, we employed the NASA-TLX workload index~\cite{hart1988development} on a 7-point scale.

Additionally, previous research in design has identified six core dimensions~\cite{sas2014generating} for assessing the communication quality of design implications: validity, generalizability, originality, generativity, inspirability, and actionability. To determine if the quality of communicated design insights differed when using \system{} compared to not using it, we included these six dimensions as survey questions, along with the original definitions from the literature, to ensure a shared understanding. Lastly, to measure the effect of translational processes, we added relevance as an additional measure, measuring the relevance of communicated insights to their design task.

\subsubsection{Analysis}

To analyze the quantitative results (\ie{}, survey responses, behavioral data from usage logs), we initially checked and confirmed that every measure met the normality assumption using the Shapiro-Wilk test. After confirming normality, we conducted a paired t-test to assess the differences in these measures between the two conditions (\system{} versus baseline).

For the qualitative results, we conducted a thematic analysis~\cite{braun2006using} with a bottom-up approach. First, two authors manually read through the transcripts and identified the initial set of themes individually. Then, they regularly met to discuss and refine the themes, which were repeated for three rounds. As a result, the authors identified the themes and corresponding quotes as detailed in \S\ref{sec:results}.

\subsection{Results}\label{sec:results}

We organize our findings into five themes: (1) overall perceptions of \system{}, (2) designer-centered retrieval, (3) clustering for managing and validating insights, (4) step-by-step translation, and (5) visualizing action items for fast interpretation and application.

\subsubsection{Overall perception of \system{}}

Participants thought positively about using \system{} to support their UI mockup iterations. When comparing between \system{} and baseline conditions, 11 participants preferred using the plugin to support their design iteration. Only one participant mentioned that their preference depends on the context, but indicated that they would only prefer to manually gather design insights from research papers if they had unlimited time to conduct iterative paper searches, admitting that this timeframe is practically impossible.

\begin{figure*}[t!]
    \centering
    \includegraphics[width=\linewidth]{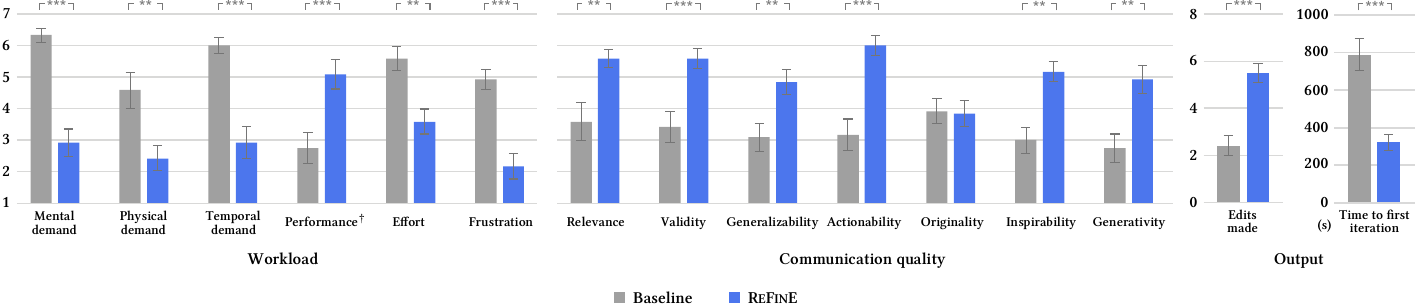}
    \caption{Perceived workload, communication quality, and the design output metrics from our user study. Participants in the \system{} condition reported significantly reduced workload compared to the baseline condition, while finding the communicated design insights to be significantly improved. Additionally, participants made significantly more edits interacting with \system{}, while reaching iteration more quickly. Error bars indicate standard errors. (**: $p<.01$, ***: $p<.001$; †: reverse item)}
    \label{fig:stats}
\end{figure*}

As shown in \autoref{fig:stats}, participants found the \system{}-driven iteration to be significantly less burdensome than the baseline condition. They found that interacting with \system{} to guide their design iterations was significantly less mentally ($M=2.92$, $SD=1.51$ vs. $M=6.33$, $SD=0.78$; $t=6.84$, $p<.001$), temporally ($M=2.92$, $SD=1.73$ vs. $M=6.00$, $SD=0.85$; $t=5.41$, $p<.001$), and physically ($M=2.42$, $SD=1.38$ vs. $M=4.58$, $SD=1.98$; $t=3.86$, $p<.01$) demanding, compared to the baseline condition. They also perceived the \system{} support as leading to better performance ($M=5.08$, $SD=1.62$ vs. $M=2.75$, $SD=1.66$; $t=4.84$, $p<.001$), while requiring less effort ($M=3.58$, $SD=1.38$ vs. $M=5.58$, $SD=1.31$; $t=3.13$, $p<.01$) and resulting in lower frustration ($M=2.17$, $SD=1.40$ vs. $M=4.92$, $SD=1.08$; $t=5.40$, $p<.001$).

Our findings also indicate that the design insights communicated by \system{} exhibited enhanced communication qualities. Participants rated \system{} as providing insights that were more relevant ($M=5.58$, $SD=0.99$ vs. $M=3.58$, $SD=2.11$; $t=2.97$, $p<.01$), generative ($M=4.92$, $SD=1.51$ vs. $M=2.75$, $SD=1.54$; $t=3.03$, $p<.01$), inspirational ($M=5.17$, $SD=1.11$ vs. $M=3.00$, $SD=1.41$; $t=3.68$, $p<.01$), actionable ($M=6.00$, $SD=1.13$ vs. $M=3.17$, $SD=1.75$; $t=4.53$, $p<.001$), valid ($M=5.58$, $SD=1.08$ vs. $M=3.42$, $SD=1.68$; $t=4.73$, $p<.001$), and generalizable ($M=4.83$, $SD=1.40$ vs. $M=3.08$, $SD=1.56$; $t=2.96$, $p<.01$) compared to the baseline. Our translational process did not compromise perceived originality ($M=3.83$, $SD=1.40$ vs. $M=3.92$, $SD=1.38$; $t=0.15$, $p=.44$).

Supporting these perceived improvements in workload and communication qualities, participants made significantly more edits with \system{} within the same timeframe. On average, participants made 5.5 design edits ($SD=1.4$) using \system{}, which was significantly higher than without the support, where they made 2.4 edits on average ($SD=1.4$; $t=6.59$, $p<.001$).

Participants also found \system{}'s generation speed sufficiently fast for their usage scenario. Out of all participants, ten responded that latency in \system{}'s content generation did not affect their design process. Two others noted that they `at least did notice' the delay, although they both reported that it did not significantly impact their design iteration. The system displayed a circular progress bar while loading; two participants suggested that providing detailed text to explain what is being generated, along with indicating the actual progress, could help mitigate the perceived effects of latency even further.

To this end, participants expected the \system{} system to be highly beneficial for their future design work, as it provided scholarly evidence to support their iterations---resources they value but often struggle to utilize~\cite{colusso2017translational}. They illustrated the excessive iterations they typically face during the prototyping stage and the difficulty of looking up multiple sources to inform their edits, highlighting the potential of \system{} in their workflows. Consequently, 10 participants expressed willingness to use the plugin in their future design process, and three interviewees directly inquired about its public release: \textit{``When are you gonna publish this plugin? I would really love to use this in my designing.''}~(P1)

Specifically, participants discussed several real-world scenarios where \system{} could be particularly valuable. Four participants envisioned using it in corporate design teams, where justifying design iterations with scholarly-backed rationale is highly appreciated, viewing \system{} as a powerful support for enhancing design iterations with well-founded insights: \textit{``I would love to use that at work because sometimes being a designer is about presenting the design to all the stakeholders, especially the PMs. So it would be nice to have evidence to back the design decisions up.''}~(P3) Similarly, three participants saw its potential in classroom design projects, where students must justify their decisions with evidence. They noted that \system{} could serve as an educational tool, fostering learning through hands-on engagement with insight-evidence pairs, ultimately enhancing both the quality of their work and their ability to articulate design rationale: \textit{``When I need to make a digital product, and if I need to make a report, it would be really helpful, because every school assignment requires reference and asks me to think through design rationale about my design iterations.''}~(P11)

\subsubsection{Designer-centered retrieval facilitated the intentional discovery of scholarly insights}\label{sec:qual-1}

Of all, 9 responded that the most painful process during the baseline condition was searching for the papers that could potentially benefit their design iteration. More specifically, they responded that they had to try out multiple queries regarding the designs, as it was difficult to find the right query for retrieving papers, and even then, they could not consistently find inspiring papers.  Supporting this, participants during the baseline condition spent the first 13m 8s on average ($SD =$ 4m 58s) out of the 25-minute task time navigating scholarly resources before beginning their initial design iteration, frequently refining their queries to gather relevant insights. This was significantly longer than in the \system{} condition, where participants reached design iteration on the canvas more quickly, spending an average of 5m 23s ($SD =$ 2m 28s) before starting their initial design iteration ($t = 5.88$, $p < .001$). During this time, while they viewed 5.8 papers ($SD=4.8$) on average during the baseline condition, they only referred back to 1.4 papers ($SD=1.0$) when designing. This repetitive and time-consuming process was extremely burdensome and the overall process less effective: \textit{``The most frustrating thing (in the baseline condition) was that, I didn't know the right keywords for finding the research paper. I felt like it was not working. I needed to try very hard to search for the relevant resources (...) which failed.''}~(P5) This increased difficulty in retrieving papers likely contributed to the higher perceived workloads during the baseline condition.

On the other hand, participants noted that \system{}'s design-centric indexing and retrieval enabled them to discover more relevant and applicable insights with less effort, in contrast to manually coming up with queries through a trial-and-error search process. First, every participant agreed that the design contexts extracted by \system{} were accurately elicited for retrieval: \textit{``I found they (design context dimensions elicited by \system{}) were really perfect ones.''}~(P12) With its accuracy, \system{}'s automatic retrieval was reported to facilitate more intentional exploration, while reducing frustration and cognitive load compared to conventional retrieval methods: \textit{``Having a plugin that can automatically collect all the papers that are related to what you're doing is really efficient.''}~(P9)

Envisioning the use of \system{} in their own projects, two participants expressed a desire to expand the design context dimensions by incorporating their own design goals---ideas that are not yet visible in their mockups but exist in their minds or elsewhere in their workflow. Currently, \system{} lets designers refine the design context dimensions by iterating on the detected dimensions after the system elicits them. However, since these dimensions are derived based on what is already visible in the design, participants wished for more control over the outputs by integrating what remains invisible, such as undocumented thoughts or goals recorded elsewhere: \textit{``(what if) this is the direction that I'm pushing for now and I'm just trying to find the evidence (...) possibly adding a way to provide my own goal to that plugin would be nice.''}~(P3)

\subsubsection{Clustering helped them to avoid information overload while improving validity}

Participants in our study found the design implications to be well-clustered. Among the seven participants who commented on clustering, six stated that the design implications were effectively grouped and accurately reflected the cluster content, and the other participant mentioned that a more thorough evaluation would be possible if they had access to the original papers: \textit{``The clusters were accurate and related to the contents (implications) they had (...) the titles clusters had were really relevant to the sources.''}~(P2)

Our survey indicated that participants considered the design insights communicated by \system{} to be significantly more valid than those they identified in the baseline condition, while also requiring significantly lower temporal demand. Participants mentioned that multiple implications provided within the cluster reinforced the cluster's core message, while enhancing its validity and eliminating the need for processing and verifying multiple papers: \textit{``Cluster is very straightforward to me (...) the suggestion is trusted, following its research paper resources. Also very easy to read.''}~(P5)

At the same time, our interview highlighted the potential need to allow designers to organize action items by each screen in their design workspace. Participants emphasized that screens serve as their primary focus, suggesting that action items derived from cluster contents should be grouped by screen first. This approach would allow designers to assess validity while maintaining alignment with their original workflow of interacting with screens: \textit{``So, for example, click a page, and then I see the action items on that page, and then maybe it has guidelines like `add a search bar here because that will (...)' to simplify the user flow.''}~(P10)

\subsubsection{Step-by-step translation helped them to broaden the scope of relevant insights without losing the originality of sources.}

Our survey suggests that, with \system{}, participants found the communicated design insights from scholarly papers more relevant for their work. In \S\ref{sec:qual-1}, we discussed how that is partly due to participants being able to find more relevant papers (\ie{}, supporting retrieval). However, at the same time, our interview results also suggest that the difference in relevance rating is partly due to the research translation that occurred. Of all, 10 noted that the translated insights from \system{} broadened their understanding of what constitutes `applicable design insights' from research papers. Participants reported a tendency during the baseline condition to fixate on finding a perfectly matching paper, which constrained their perception of applicable insights: \textit{``I was kind of being oriented by the paper directions, not being led by my own thoughts. I don't like that process, and I would never do that again.''}~(P11) However, with \system{}'s translation support, they were able to explore a wider range of relevant insights that they would have overlooked in traditional paper querying contexts: \textit{``The use cases for me would be that it inspires me to consider aspects I might not think about otherwise if I weren't using the tool.''}~(P4) This reinforces previous findings that translating design insights from research papers expands the scope of relevant scholarly insights~\cite{shin2025what}.

In this process, the step-by-step guidance provided by \system{} played a crucial role in helping participants assess the validity of the translated insights without losing the originality of the sources. Beginning with a comparison and contrast, they could gain an overview of these implications before seeing how they could be applied to their specific contexts. This structured process enabled them to quickly and effectively evaluate the relevance and applicability of the translated insights: \textit{``It told me that some of the papers were relevant, then they would explain that it's because those papers also talk about travel (as a domain), and that allowed me to trust it. And then also with the plugin they had a summary. And they're like, oh, this is how you can inspire your design. That was something that the papers I reviewed (in the baseline condition) didn't have.''}~(P10)

\begin{figure}[t!]
    \centering
    \includegraphics[width=.8\linewidth]{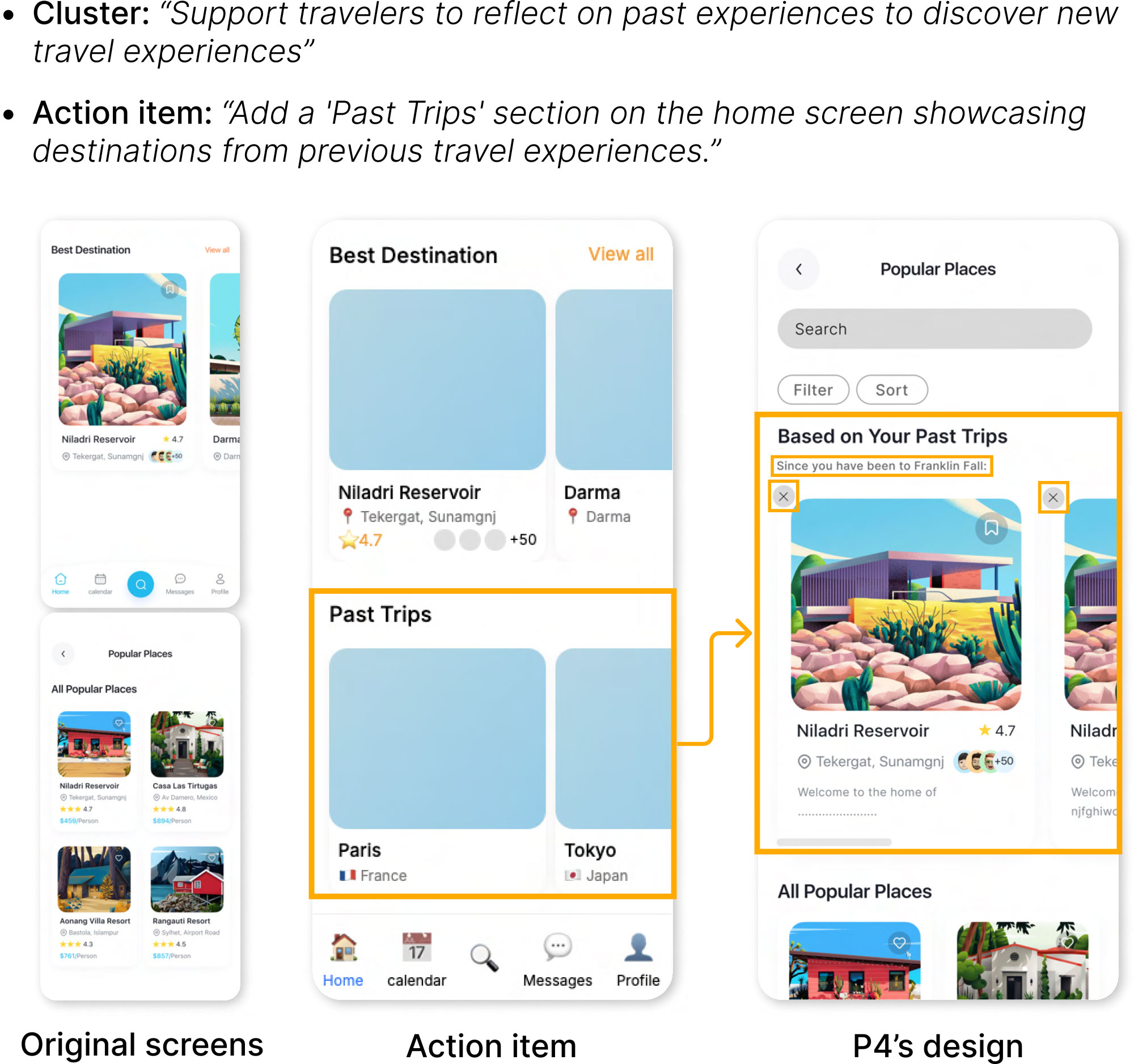}
    \caption{P4's example of implementing an action item. After recognizing its potential usefulness, P4 creatively adapted the action item to a different screen, redesigning it with alternative UI components.}
    \label{fig:action-item-examples}
\end{figure}

\subsubsection{Visualizing action items improved designers' ability to make informed design decisions quickly and enhanced learnability}\label{sec:user-study:visualizing-action-items}

Describing themselves as `visual learners,' participants emphasized the significance of visual illustrations in understanding design insights. Before the study began, they expressed that the text-heavy nature of papers made it challenging to derive inspiration from written content, which diminished the papers' utility---which aligns with previous findings in translational science for design that noted the difficulties associated with consuming text-heavy papers~\cite{colusso2017translational}. Similarly, participants during the study reported challenges in quickly grasping the design implications presented in the text of the papers: \textit{``I feel like that's just too much text (in the paper) and there are no highlights relating to the design suggestions.''}~(P3)

Our survey showed that participants rated the actionability and validity of design insights significantly higher when using \system{} compared to the baseline, with lower temporal demand and more design edits made within the same timeframe. Our interviews showed that visualizing action items allowed participants to quickly understand the required actions, streamlining their process of validating the insights and minimizing the time needed to integrate them into their designs. All participants reported the visual representations of the action items to be extremely useful, as they not only enabled the quick application of insights to their designs but also facilitated a better understanding of the messages conveyed. By rapidly grasping the insights through the visuals, participants could assess their usefulness based on their design experience and the relevance of surrounding content (\eg{}, cluster contents, sources), which streamlined the process while maintaining their validity: \textit{``I think that (visualization) is the really intuitive and direct way of understanding of explanation on the rationale. I had a pretty good understanding when I just saw the visuals.''}~(P11)

With a deeper understanding, we observed that the visualization also facilitated externalization, allowing participants to use it as a learning tool to translate insights into their own understanding. As shown in \autoref{fig:action-item-examples}, the enhanced comprehension of action items through visualization enabled participants to externalize these items and proactively design in ways they deemed improved. This demonstrated that visualization not only reinforced understanding but also empowered participants to take an active role in refining and applying design concepts.

In this process, every participant unanimously reported that the minor discrepancies in the reconstructed mockups did not affect their comprehension of the action items. They indicated that they were more focused on the overall `semantics' conveyed by the visualization rather than minor visual details. Here, the presence of a toggle to turn the visual change on or off was reported to significantly assist in their prompt understanding of the modifications: \textit{``It had all the same components, like same features, buttons, and all that. It was just different and kind of different in design. For me, I see all of these as components, and the design was up to me.''}~(P9)

\subsubsection{Enhanced scannability and optional presentation of contents to better help the navigation and validation of design insights.}

Although our quantitative results indicate that participants found the use of \system{} to be significantly less effortful compared to the baseline, we identified opportunities to further reduce effort, such as incorporating a more catchy summary of clusters to streamline the process of discovering insights. For instance, upon encountering a cluster about the importance of improving financial literacy for mobile banking security, P9 suggested adding hashtags like `financial literacy' to facilitate prompt understanding of what the cluster is talking about: \textit{``I think maybe it could be better to have keywords, like `financial literacy' (...) just showing those keywords would make it easier to navigate.''}~(P9) Similarly, to further improve the scannability of the contents, participants highlighted the potential need for highlighting the subset of contents, such as boldfacing the important keywords: \textit{``If only the core keywords can be highlighted, or bold, maybe it's helpful for quickly scanning the clusters.''}~(P4)

Additionally, participants acknowledged that supporting evidence (\eg{}, sources) contributed to originality and validity; yet, after reviewing these and finding the content trustworthy, they noted that sources were not their primary focus when reading in their design scenario. As a result, they felt these elements occupied excessive space and preferred to hide them by default, opting to review them only when insights contradicted their expectations and required validation. Consequently, they suggested making these sections hidden by default and expandable upon request: \textit{``So maybe we can hide that, and then you can open it only when you are interested in learning more about the sources.''}~(P12)
\section{Discussion \& Future Work}

\subsection{Supporting Design Insight Consumption Through Visualization}

In this paper, we introduced \system{}, a novel system powered by multimodal LLMs that streamlines the consumption of design implications from research papers during UI design iterations in Figma. \system{} tackles key challenges across retrieval, translation, and contextualization---common barriers to applying design insights, as identified in prior research in translational science for design. Through our technical evaluations and user study, we demonstrated the reliability of \system{}'s components and explored how designers perceive its impact during real design iterations.

One key implication of our study is that supporting the visualization of scholarly insights into their designs not only facilitates their quick integration into workflows but also significantly enhances the understanding and learnability of these insights. Designers, who are heavily trained to engage with visuals, often find text-heavy content overwhelming~\cite{park2024we, colusso2017translational, howell2021exploring}, as supported by our user study revealing their challenges in consuming full-text papers and lengthy summaries. With design implications presented in a visual format, they could quickly grasp the core messages of each implication cluster and more easily incorporate them into their design iteration. Still, our system currently limits visualization to static design components that can be integrated without adding extra screens or interactivity. Having demonstrated the efficacy of our approach, future research could explore ways to support dynamic elements or those requiring additional screens, expanding the applicability of our approach.

\subsection{Implications of AI-Mediated Design Support}

While user outcomes were positive, our findings also suggest possible unintended effects and limitations of AI-mediated design iteration. For instance, we observed a pattern in which action items generated by \system{} tended to be additive---introducing new UI elements or affordances (\eg{}, examples in \autoref{fig:proof-by-demonstration})---rather than subtractive or simplifying existing designs. We hypothesize this bias may stem from several sources, including conventions in how design implications are presented in research literature, our system design, or broader tendencies of LLMs to favor additive transformations. Importantly, it remains unclear whether additive recommendations are the most appropriate or desirable in these contexts, as reducing or simplifying UI elements can also enhance usability and accessibility, or reduce cognitive load. This tendency underscores the importance of real-world evaluation. No users in our study explicitly identified this point as a negative, potentially because they retained agency over the ultimate design choices, though future work should investigate the impacts of this additive bias on designer decision-making.

Another finding from our study is that, when designers were asked to search for papers directly in the baseline condition, most designers relied on narrow, domain-specific keyword searches in the paper repository, limiting their exposure to relevant papers outside those exact terms. Our system operates within the Figma canvas to help match users with papers along dimensions critical to design relevance and uses that structured understanding, augmented by translation, to retrieve papers that may not match the original keywords but still offer valuable insights. This helped designers move beyond keyword constraints and discover findings more aligned with their design needs. In this process, the informational content within each insight cluster served as an explainability layer, guiding designers on how to adapt insights from imperfectly matching papers. These results highlight the potential for a more expansive notion of relevance when retrieving from paper repositories. At the same time, our findings identify a need for more concise cluster summaries; future work could explore presentation methods for highlighting key takeaways to enhance navigation and scannability.

\subsection{Rethinking Retrieval and Expanding Literature Scope}

While our system demonstrated its effectiveness in supporting design iteration using a limited set of papers (\ie{}, \proceedings{} proceedings), we see potential in increasing the scope---both extending (\ie{}, covering multiple-year proceedings) and expanding (\ie{}, diversifying the type of proceedings) the sources of literature. First, extending the literature to multiple proceedings could increase the granularity of topical coverage, improving the chances of finding more relevant literature. Second, although our studies utilized CHI proceedings due to its broad coverage, the focus of more domain- and topically-specialized HCI conferences (\eg{}, DIS, UIST, ASSETS, MobileHCI) could provide designers with a filtering mechanism to retrieve papers more specialized to their designs. As such, incorporating the unique focus of each venue as an additional dimension could enhance \system{}'s ability to recommend design insights tailored to specific designer needs, yet it would be necessary to understand if the papers retrieved from these added conferences have sufficient diversity and relevance. Beyond HCI, future work could explore incorporating publications from other applied disciplines (\eg{}, psychology, communication), which often produce research relevant to design practices. \system{} will likely need to be modified to support the translation of papers from these scholarly communities with different practices for communicating practical implications.

In this paper, we presented an evaluation of \system{} in a mobile UI design setting, building on prior works that focused on mobile UIs as a testbed (\eg{},~\cite{duan2024generating, wang2023enabling}). However, our system should be readily adaptable to other types of interfaces. For example, Figma is widely used by designers to create UIs on other modalities (\eg{}, web\footnote{https://www.figma.com/community/design-tutorials/web-design}, smart watch\footnote{https://www.figma.com/community/search?resource\_type=mixed\&query=watch}), and \system{} can dynamically detect the design modality through the elicitation of design context, enabling the retrieval of insights tailored to the corresponding modality. Beyond Figma, \system{} can also be integrated with commercial web editors that support WYSIWYG and HTML exports, enabling faster and more accurate mockup visualizations by directly utilizing their native HTML exports, thereby eliminating the need for mockup reconstruction.

\subsection{Future Work}

Our findings show that \system{} supports rather than replaces designer expertise. Designers in our user study consistently exercised independent judgment by filtering, modifying, or selectively integrating these suggestions instead of adopting them uncritically (\S\ref{sec:user-study:visualizing-action-items}). This highlights the interpretive and situated nature of design work, where design insights are reconciled with contextual constraints, aesthetic goals, and tacit knowledge. Moving beyond, future translational systems could more explicitly leverage designer expertise by supporting iterative feedback loops between designers and the system; for example, enabling designers to critique, annotate, or revise suggested actions and have those responses shape subsequent recommendations. Such mechanisms would more robustly model and adapt to designer intent over time.

Furthermore, when \system{} encounters conflicting or contradictory design insights within a cluster, it currently resolves this by instructing the LLM to prioritize the insight most aligned with the designer's context. While this streamlines the generation of actionable items, it may inadvertently obscure valuable design tensions. Future translational systems could explore surfacing these contradictions explicitly to the designer, such as by presenting alternative design actions or highlighting the trade-offs between conflicting guidelines, thereby empowering designers to navigate these complexities using their own expertise rather than relying on automated resolution.

Lastly, participants envisioned various potential use cases within their real-world design workflows, such as applying \system{} in corporate settings or class projects, while at the same time revealing the need to extend the design contexts extracted by \system{} beyond visible elements to include aspects embedded in their thoughts or workflows. To better support these scenarios, we propose that future translational systems for UI design iterations could leverage documents from existing workflows and extract implicit design goals, as these contexts are often guided by well-documented guidelines or objectives (\eg{}, marketing briefs, syllabi) that may not be immediately apparent in a design mockup. By incorporating these additional dimensions for characterizing a design context, the system could further enrich results by aligning them with organizational goals. Similarly, future work can explore methods to compare \system{}'s recommendations against established sources of design knowledge, such as design textbooks, curated professional guidelines, and tutorials. Such comparisons could help assess the alignment, complementarity, or potential gaps between research-derived insights and practitioner-oriented resources, and provide a more holistic understanding of how different knowledge sources can support design iteration.
\section{Conclusion}

In this work, we introduce \system{}, a system designed to seamlessly integrate scholarly design implications into the UI mockup iteration process. \system{} automates the process of retrieving, translating, and contextualizing research insights from HCI papers, enabling designers to incorporate evidence-informed knowledge directly within their design workspace (\ie{}, Figma). Our comprehensive evaluations, including technical evaluations and a user study with designers, demonstrate the reliability of \system{} components and the system's ability to reduce the burden on designers, improve the communication of scholarly insights, and streamline design iteration. Our results highlight \system{}'s potential to bridge the gap between academic research and practical design, providing insights for developing tools that support evidence-informed design workflows.

\begin{acks}
We thank Joanne Ma, Daniel Lee, Soomin Kim, and the members of the UW Prosocial Computing Group \& UW LARCH Lab for their valuable feedback on this paper. This project is supported by the Microsoft AI and the New Future of Work Award and the Google PaliGemma Academic Program GCP Credit Award.
\end{acks}

\bibliographystyle{ACM-Reference-Format}
\bibliography{100-bibliography}

\appendix
\newpage
\onecolumn
\section{Technical Implementation Details}\label{app:algorithms}

\subsection{Retrieval Algorithm Formalization}

As described in the main text, we represent the user's design mockup as a query vector: \(m = [m_1, \dots, m_6] \), where each $m_i$ corresponds to one of the six extracted design context dimensions. Similarly, the paper vector index consists of embedding arrays representing the six dimensions of each paper in the repository:
\[
P = \left[ [p_{1_1}, \dots, p_{1_6}], [p_{2_1}, \dots, p_{2_6}], \dots \right]
\]

To ensure robust retrieval, our system prioritizes dimensions that are present in both the design mockup and the paper (\ie{}, valid dimensions). We compute a summed embedding over all valid dimensions for the mockup representation as $S_m = \sum_{i \in \text{valid}} m_i$ and the corresponding dimensions of each paper in our retrieval index as $S_{p_i} = \sum_{j \in \text{valid}} p_{i_j}$. \system{} retrieves papers based on the cosine similarity between \( S_m \) and each \( S_{p_i} \) as:
\[
sim(S_m, S_{p_i}) = \frac{S_m \cdot S_{p_i}}{\|S_m\| \|S_{p_i}\|} \quad \forall p_i \in P
\]

\subsection{Clustering Optimization Formalization}

We employ hierarchical clustering over the embeddings of the retrieved design implications. These embeddings are represented as \( \mathcal{D} = \{d_1, d_2, \dots\} \), where each design implication \( d_i \) has an associated embedding vector \( \mathbf{d}_i \in \mathbb{R}^n \). To determine the optimal number of clusters, the system tests various values of \( n_{\text{clusters}} \in \{2, 3, \dots, n_{\max}\} \) and computes the silhouette score~\cite{rousseeuw1987silhouettes} for each clustering.  

For each data point \(d_i \in \mathcal{D}\), the silhouette score is defined as:
\[
s_i = \frac{b_i - a_i}{\max(a_i, b_i)}
\]
where \(a_i\) is the average distance from \(d_i\) to other points in the same cluster, and \(b_i\) is the average distance from \(d_i\) to points in the nearest cluster that \(d_i\) is not part of. The silhouette score for a clustering \(C\) is then the mean over all points:
\[
S(C) = \frac{1}{|\mathcal{D}|} \sum_{i=1}^{|\mathcal{D}|} s_i
\]

The optimal number of clusters, \(n_{\text{best}}\), is the one that maximizes the mean silhouette score:
\[
n_{\text{best}} = \operatorname*{argmax}_{n_{\text{clusters}}} S(C_n)
\]
Using \(n_{\max}=10\), the implications are grouped into \(n_{\text{best}}\) clusters. Let \(C_k\) represent the set of design implications in cluster \(k\). These groups are stored as:
\[
\mathcal{G} = \{ C_1, C_2, \dots, C_{n_{\text{best}}} \}
\]
The final output for each cluster \(C_k\) is:
\[
\mathcal{R}_k = \left\{ \left( \mathcal{I}_k, \mathcal{S}_k \right) \mid C_k = \{d_1, d_2, \dots\} \right\} \quad \forall C_k \in \mathcal{G}
\]
where \(\mathcal{I}_k\) is the list of original implications and \(\mathcal{S}_k\) denotes the set of translated insights derived from each cluster.

\section{Study Details}\label{apdx:study-details}

\subsection{Technical Evaluation}
\begin{table}[H]
    \caption{Comparison of VLMs for HTML reconstruction using image inputs (metrics are computed per-screen $\pm$ stdev)}
    \footnotesize
    \begin{tabular}{lcc}
    \toprule
    \textbf{Model} & \textbf{Visual similarity} & \textbf{Latency (seconds)} \\
    \midrule
    Gemini 2.0 Flash~\cite{gemini} & \(0.7875 \pm 0.1437\) & \(10.794 \pm 3.021\) \\
    Claude 3.5 Sonnet~\cite{claude} & \(0.7963 \pm 0.1590\) & \(17.006 \pm 4.914\) \\
    GPT-4o~\cite{gpt4o} & \(0.7852 \pm 0.1330\) & \(11.060 \pm 3.382\) \\
    \bottomrule
    \end{tabular}
    \label{tab:model-comparison}
    \Description{}
\end{table}
\begin{table}[H]
    \caption{Comparison of different input modalities on HTML reconstruction (base model is Gemini 2.0 Flash, metrics are computed per screen $\pm$ stdev)}
    \footnotesize
    \begin{tabular}{lcc}
    \toprule
    \textbf{Input modality} & \textbf{Visual similarity} & \textbf{Latency (seconds)} \\
    \midrule
    Image & \(0.7875 \pm 0.1437\) & \(10.794 \pm 3.021\) \\
    JSON & \(0.7661 \pm 0.1314\) & \(15.432 \pm 7.047\) \\
    Image + JSON & \(0.7711 \pm 0.1411\) & \(10.630 \pm 2.427\) \\
    \bottomrule
    \end{tabular}
    \label{tab:modality-comparison}
    \Description{}
\end{table}

\begin{table}[H]
    \caption{Comparison of methods for generating visualizations of action items applied to designs (metrics are computed per action item $\pm$ stdev)}
    \footnotesize
    \begin{tabular}{lcc}
    \toprule
    \textbf{Method} & \textbf{Accuracy} & \textbf{Latency (seconds)} \\
    \midrule
    Complete HTML regeneration & \(96.0\%\) & \(10.248 \pm 3.488\) \\
    Edit-only generation & \(95.3\%\) & \(2.913 \pm 1.215\) \\
    \bottomrule
    \end{tabular}
    \label{tab:preview-comparison}
    \Description{}
\end{table}
\begin{table}[H]
\centering
\footnotesize
\caption{Examples of relevance ratings assigned by annotators when evaluating the connection between design implications and action items}
\begin{tabularx}{\textwidth}{
>{\raggedright\arraybackslash}p{2.5cm} 
>{\raggedright\arraybackslash}p{3.5cm} 
>{\raggedright\arraybackslash}p{4.5cm} 
>{\raggedright\arraybackslash}X
}
\toprule
\textbf{Relevance rating} & \textbf{Design implication} & \textbf{Action item} & \textbf{Annotator's explanation} \\
\midrule
5 - Highly relevant & Reduce user burden by minimizing text input in form-heavy tasks & Implement a one-click autofill feature that uses stored preferences for all checkout fields & Directly addresses and solves the burden \\
\addlinespace
4 - Substantially relevant & Make it easier for users to navigate dense information in reports & Add a ‘Back to Top’ button on the settings screen & Helps with navigation, but not specific to navigating documents \\
\addlinespace
3 - Moderately relevant & Striking the balance between insufficient and overwhelming transparency will enable users to trust the system & Add a ‘Report’ button to the comment section to enhance user feedback and platform moderation & May support trust, but unclear link to transparency or balance \\
\addlinespace
2 - Slightly relevant & Allow users to control the visibility of their personal data in shared platforms & Let users toggle switches in the profile settings to show/hide recent activity & Some connection to visibility, but lacks clarity on shared context \\
\addlinespace
1 - Not relevant & Help users build trust in financial transactions through transparency & Display an autoplay video on the app with product promotions & Unrelated to trust or transparency \\
\bottomrule
\end{tabularx}
\label{tab:relevance}
\Description{This table illustrates examples of relevance ratings from 1 (Not relevant) to 5 (Highly relevant), assigned by annotators assessing how well action items align with design implications. Each row includes a rating, a design implication, a corresponding action item, and a brief explanation justifying the assigned relevance.}
\end{table}

\subsection{User Study}\label{apdx:study-details-user-study}
\begin{figure}[h!]
    \centering
    \includegraphics[width=.4\linewidth]{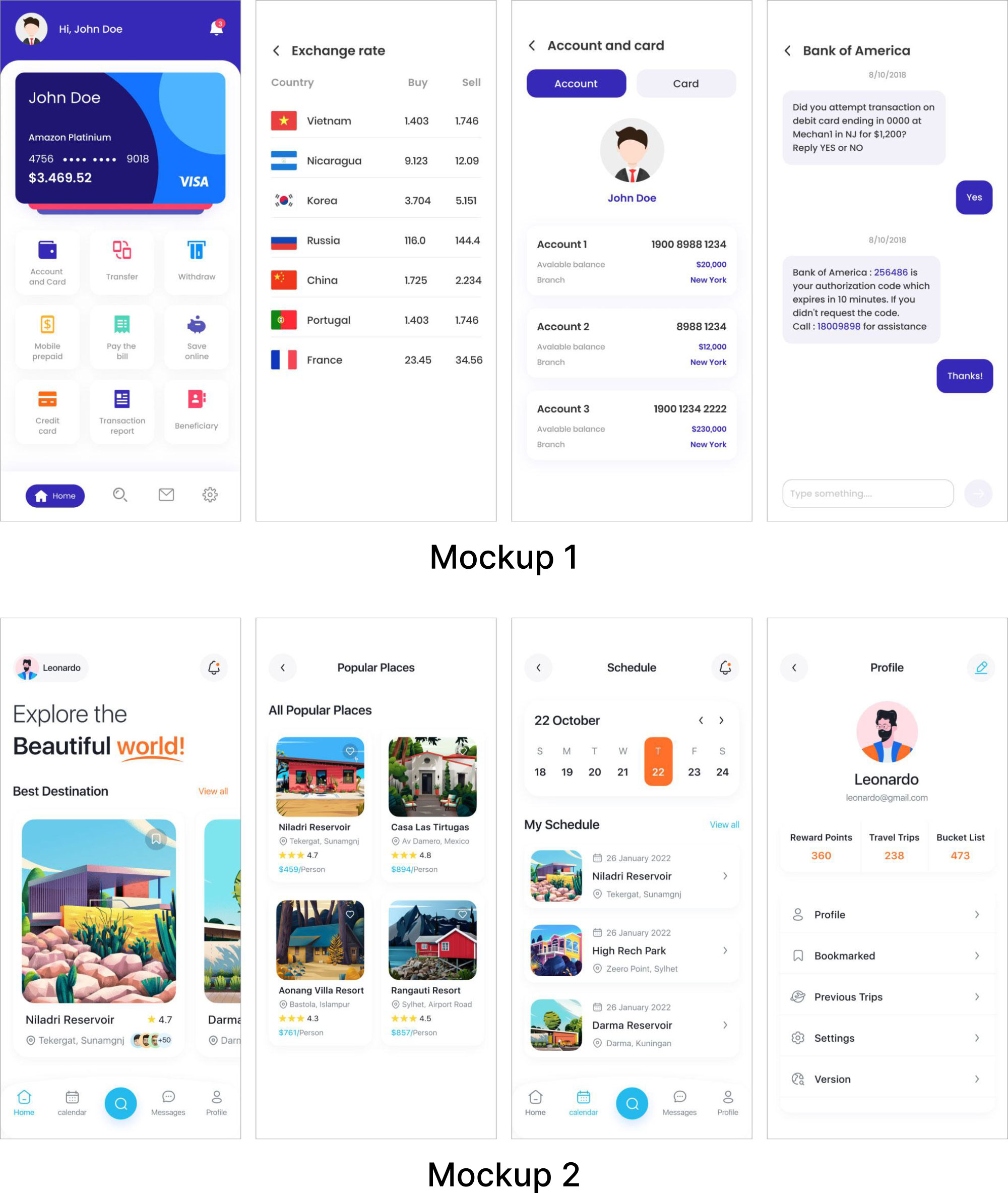}
    \caption{Prototype mockups used in our user study}
    \label{fig:study-mockup}
\end{figure}

\end{document}